\documentclass[twocolumn,letter]{jpsj3}

\usepackage{color}
\usepackage{bm}

\title{%
Weak Topological Insulators with Step Edges:\\
Subband Engineering and Its Effect on Electron Transport
}

\author{%
Takashi Arita and
Yositake Takane\thanks{takane@hiroshima-u.ac.jp}
}

\inst{%
Department of Quantum Matter, Graduate School of Advanced Sciences of Matter,\\
Hiroshima University, Higashihiroshima, Hiroshima 739-8530, Japan
}

\recdate{ \hspace{50mm} }

\abst{%
A three-dimensional weak topological insulator (WTI) can be regarded as
stacked layers of two-dimensional quantum spin-Hall insulators,
each of which accommodates a one-dimensional helical edge mode.
Massless Dirac electrons emerge on a side surface of WTIs
as a consequence of the hybridization of such helical edge modes.
We study the energy spectrum and transport of Dirac electrons
on a side surface in the presence of step edges,
which significantly modify the way of hybridization.
It is shown that pseudo-helical modes with a nearly gapless linear dispersion
can be created by manipulating step edges in a certain manner.
We numerically calculate the average conductance of weakly disordered WTIs
and show that it is markedly enhanced by pseudo-helical modes.
}

\begin{document}
\sloppy
\maketitle

Three-dimensional weak topological insulators (WTIs)
can be regarded as stacked layers of
two-dimensional quantum spin-Hall insulators (QSHIs).~\cite{fu,moore,roy}
Each QSHI layer possesses gapless excitations only along its edge
in the form of a one-dimensional (1D) gapless helical mode
consisting of a pair of counterpropagating channels.\cite{kane,bernevig}
As a result of the hybridization of 1D helical edge modes,
low-energy electrons obeying a massless Dirac equation
appear on a side surface of WTIs.
Such electrons are referred to as Dirac electrons.
Their unusual characters have been extensively studied
so far.~\cite{ran,imura1,ringel,mong,liu1,imura2,yoshimura,kobayashi1,
morimoto,obuse,takane1,arita,takane2,matsumoto,kobayashi2}
Several materials have been proposed or confirmed
as WTIs.~\cite{yan,rasche,tang,g-yang,pauly,zhou}

Let us consider electron transport on a side surface of WTIs
consisting of $M$ layers of QSHIs.
A 1D helical edge mode of an isolated QSHI serves as
a perfectly conducting channel (PCC) without backscattering
even in the presence of disorder.
However, once $M$ helical modes are coupled on a side surface,
the resulting system possesses at most one PCC depending
on the parity of $M$.~\cite{ringel,ando1,takane3,takane4,takane5,ando2}
If $M$ is even, all subbands acquire a finite-size gap, leading to no PCC.
If $M$ is odd, one subband remains gapless and this induces one PCC.
Note that, if $M$ layers were completely decoupled,
there should be $M$ gapless helical modes.
This indicates that the number of gapless modes,
or equivalently the number of PCCs, is reduced from $M$ to at most one
owing to the hybridization.
Conversely, this suggests that we can modify
the energy spectrum and therefore the electron transport
if the way of hybridization can be manipulated.

To manipulate the way of hybridization, we propose to use straight step edges
longitudinally arranged on a side surface of WTIs (see Fig.~1).~\cite{arita}
The coupling of neighboring 1D helical edge modes is weakened
if such a step edge is present between them,
and the coupling strength becomes small with increasing width of the step.
In the presence of $m$ step edges with a sufficiently large width,
the system is completely separated into $m+1$ subsystems,
as far as Dirac electrons are concerned, and each subsystem possesses
one gapless mode if it consists of an odd number of layers.
Hence, we can create more than one gapless mode
by arranging step edges with a large width.
If the width is not so large,
such modes are no longer exactly gapless owing to the hybridization,
but the corresponding excitation gap should be much smaller than
that in the absence of step edges.
Let us refer to such weakly gapped modes as pseudo-helical modes.
They should significantly increase the conductance
of disordered WTIs.
%%%%%%%%%%%%%%%%%%
\begin{figure}[tbp]
\begin{center}
\includegraphics[height=2.4cm]{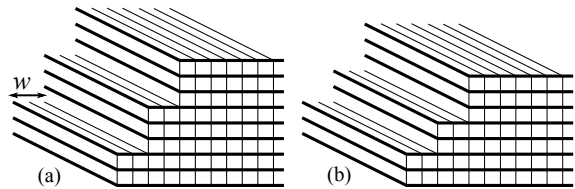}
\end{center}
\caption{WTI samples consisting of (a) nine QSHI layers
and (b) eight QSHI layers with two straight step edges
of width $w = 2$ in units of the lattice constant $a$.
Thick lines denote QSHI layers.
}
\end{figure}
%%%%%%%%%%%%%%%%%%

The simplest prescription to induce pseudo-helical modes
is to arrange step edges so that the system is weakly decoupled
into subsystems with an odd number of layers.
For example, the nine-layer system is decoupled by two step edges into
three subsystems, each of which consists of three layers [Fig.~1(a)].
In this arrangement, symbolically denoted by $(9 \to 3/3/3)$,
there appear one exactly gapless helical mode and two pseudo-helical modes.
We also consider a slightly modified prescription in which
step edges are arranged so that an additional subsystem
with an even number of layers is placed in between
subsystems with an odd number of layers.
For example, the eight-layer system is decoupled by two step edges
into three subsystems consisting of three layers,
two layers, and three layers [Fig.~1(b)].
In this arrangement of $(8 \to 3/2/3)$, there appear two pseudo-helical modes.
An even number of layers with an excitation gap plays
the role of a buffer layer,
reducing the coupling between neighboring pseudo-helical modes across it.
That is, the coupling in this prescription
is weaker than that in the former prescription.

In this Letter, we study how straight step edges arranged on a side surface
affect the energy spectrum and electron transport.
The cases of $(9 \to 3/3/3)$ and $(8 \to 3/2/3)$ are analyzed
assuming that step edges have a common width $w$
in units of the lattice constant $a$.
Our analysis mainly relies on an effective model for Dirac electrons,
which can be derived from a tight-binding model for WTIs.
It is shown that pseudo-helical modes with a very small excitation gap
are created in the presence of step edges.
We compute the average dimensionless conductance
in the presence of weak disorder as a function of system length.
It is also shown that
the average conductance is markedly enhanced with increasing $w$.
We set $\hbar = 1$ in this letter.

%%%%%%%%%%%%%%%%%%
\begin{figure}[tbp]
\begin{center}
\includegraphics[height=2.6cm]{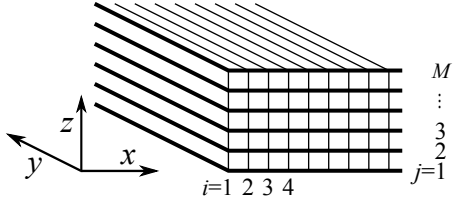}
\end{center}
\caption{WTI sample with no step edge;
it consists of $M$ layers in the $z$-direction with $M \ge j \ge 1$
while it is semi-infinite in the $x$-direction with $i \ge 1$
and is infinitely long in the $y$-direction.
}
\end{figure}
%%%%%%%%%%%%%%%%%%
Let us introduce a tight-binding model for the WTI consisting of $M$ layers
stacked in the $z$-direction (see Fig.~2), being infinitely long
in the $y$-direction and semi-infinite in the $x$-direction.
The $x$- and $z$-coordinates are discretized on a square lattice
with the lattice constant $a$, while the $y$-coordinate is continuous.
The indices $i$ and $j$ are respectively used to specify lattice sites
in the $x$- and $z$-directions.
The $y$ dependence of Dirac electron states
is characterized by the wave number $k_{y}$.
Let us introduce the four-component state vector for the $(i,j)$th site,
\begin{align}
  |i,j \rangle
  =  \left[ |i,j \rangle_{1\uparrow},
             |i,j \rangle_{2\uparrow},
             |i,j \rangle_{1\downarrow},
             |i,j \rangle_{2\downarrow}
     \right] ,
\end{align}
where $\uparrow, \downarrow$ and $1, 2$ respectively
represent the spin and orbital degrees of freedom.
The tight-binding Hamiltonian for WTIs is expressed as
$H_{\rm 3D} = H_{x}+H_{y}+H_{z}$ with~\cite{imura2,arita}
\begin{align}
   H_{x}
 & = \sum_{i=1}^{\infty}\sum_{j=1}^{M}
     \Bigl[ |i,j \rangle h_{0} \langle i,j|
            + \bigl\{ |i+1,j \rangle h_{x}^{+} \langle i,j|
                      + {\rm h.c.}
              \bigr\}
     \Bigr] ,
         \\
   H_{y}
 & = \sum_{i=1}^{\infty}\sum_{j=1}^{M}
     |i,j \rangle h_{y} \langle i,j| ,
         \\
   H_{z}
 & = \sum_{i=1}^{\infty}\sum_{j=1}^{M-1}
     \left\{ |i,j+1 \rangle h_{z}^{+} \langle i,j|
             + {\rm h.c.} \right\} .
\end{align}
Here, the $4 \times 4$ matrices are given by
\begin{align}
   h_{0}
 & = \left[ 
       \begin{array}{cc}
         \tilde{m}_{0}\tau_{z} & 0_{2 \times 2} \\
         0_{2 \times 2} & \tilde{m}_{0}\tau_{z}
       \end{array}
     \right] ,
               \\
   h_{x}^{+}
 & = \left[ 
       \begin{array}{cc}
         -m_{2\parallel}\tau_{z} & \frac{i}{2}A\tau_{x} \\
         \frac{i}{2}A\tau_{x} & -m_{2\parallel}\tau_{z}
       \end{array}
     \right] ,
               \\
   h_{y}
 & = \left[ 
       \begin{array}{cc}
         \xi(k_y)\tau_{z} & -iA(k_{y}a)\tau_{x} \\
         iA(k_{y}a)\tau_{x} & \xi(k_y)\tau_{z}
       \end{array}
     \right] ,
               \\
   h_{z}^{+}
 & = \left[ 
       \begin{array}{cc}
         -m_{2\perp}\tau_{z}+\frac{i}{2}B\tau_{x} & 0_{2 \times 2} \\
         0_{2 \times 2} & -m_{2\perp}\tau_{z}-\frac{i}{2}B\tau_{x}
       \end{array}
     \right] ,
\end{align}
where $\tilde{m}_{0} = m_{0}+2m_{2\parallel}+2m_{2\perp}$,
$\xi(k_y) = m_{2\parallel} (k_{y}a)^{2}$,
and $\tau_{x}$ and $\tau_{z}$ are respectively the $x$- and $z$-components
of the Pauli matrix representing the orbital degrees of freedom.
The weak topological phase under consideration is realized in the case of
$m_{2\parallel} > \frac{1}{4}|m_0| > m_{2\perp}
> \frac{1}{4}|m_0| - m_{2\parallel}$
with $m_{0} < 0$ and $m_{2\parallel} > 0$.~\cite{imura2}
Note that this model can be used in the presence of step edges.
In the continuum limit, $H_{3 \rm D}$
is reduced to the Hamiltonian derived in Ref.~\citen{liu2}.

We turn to the derivation of an effective model
for Dirac electrons.~\cite{arita}
The basis functions for the 1D helical edge mode arising from
the $j$th QSHI layer are obtained by solving the eigenvalue equation
$H_{x}|\psi(j)\rangle = E_{\perp}|\psi(j)\rangle$
for a fixed $j$ under an appropriate boundary condition.
The resulting functions are
$|\psi_{\pm}(j)\rangle = |\psi_{0}(j)\rangle \mib{v}_{\pm}$
with $E_{\perp} = 0$,
where $\mib{v}_{+} = {}^{\rm t} \! \left(0, -i, 1, 0 \right)/\sqrt{2}$,
$\mib{v}_{-} = {}^{\rm t} \! \left(1, 0, 0, -i \right)/\sqrt{2}$, and
\begin{align}
     \label{eq:psi_0-j}
   |\psi_{0}(j)\rangle
   = \mathcal{C} \sum_{i=1}^{\infty}
     \left(\rho_{+}^{i}-\rho_{-}^{i}\right)|i,j \rangle .
\end{align}
Here, $\mathcal{C}$ is a normalization constant, and
$\rho_{+}$ and $\rho_{-}$ satisfying $|\rho_{\pm}|<1$ are given by
\begin{align}
     \label{eq:rho-pm}
  \rho_{\pm}
  = \frac{\tilde{m}_{0}
          \pm\sqrt{\tilde{m}_{0}^{2}-4m_{2\parallel}^{2}+A^{2}}}
         {2m_{2\parallel}+A} .
\end{align}
Clearly, $|\psi_{0}(j)\rangle$ represents the penetration of
1D edge modes into the bulk.
In terms of the basis vector
$|j\rangle \equiv \left\{|\psi_{+}(j)\rangle, |\psi_{-}(j)\rangle \right\}$
representing the $j$th 1D helical edge mode,
we can express a hybridized surface state as
\begin{align}
  |\Psi\rangle
  = \sum_{j=1}^{M} |j\rangle
                   \left[ \begin{array}{c}
                            \alpha_j \\ \beta_j
                          \end{array}
                   \right] .
\end{align}
From the eigenvalue equation of
$\left(H_{y}+H_{z}\right)|\Psi\rangle = E|\Psi\rangle$,
we can find a pair of equations for $\alpha_{j}$ and $\beta_{j}$.
The effective Hamiltonian that reproduces it is given by
\begin{align}
       \label{H_2D}
   H_{\rm 2D}
 & = \sum_{j=1}^{M}
     |j\rangle \left[ \begin{array}{cc}
                          A(k_{y}a) & 0 \\
                          0 & -A(k_{y}a)
                      \end{array} \right]
     \langle j|
   \nonumber \\
 &  \hspace{-3mm}
   + \sum_{j=1}^{M-1}
     \left\{
     |j+1\rangle \left[ \begin{array}{cc}
                          0 & -\frac{1}{2}B \\
                          \frac{1}{2}B & 0
                        \end{array} \right]
     \langle j|
   + {\rm h.c.}
     \right\} .
\end{align}
This model simply describes $M$ chains of 1D helical edge modes
each of which is coupled with its nearest
neighbors.~\cite{morimoto,obuse,takane1,arita,takane2}
As is clear from Eq.~(\ref{H_2D}),
the behavior of Dirac electrons is determined by $A$ and $B$,
while the other parameters only play the role in determining
the penetration depth [see Eq.~(\ref{eq:rho-pm})].
Although the case with no step edge is implicitly assumed above,
$H_{\rm 2D}$ can be used even in the presence of straight step edges
if we appropriately reduce $B$ that
connects two neighboring helical modes across each step edge.~\cite{arita}
Let $\eta(w)$ be the factor of reduction for a step edge of width $w$.
This is determined by the overlap of the basis functions for
neighboring helical modes across a step edge as~\cite{comment1}
\begin{align}
     \label{eq:def-eta}
  \eta(w)
  = |\mathcal{C}|^{2}
    \left| \sum_{i=1}^{\infty}
           \left(\rho_{+}^{i+w}-\rho_{-}^{i+w}\right)^{*}
           \left(\rho_{+}^{i}-\rho_{-}^{i}\right)
    \right| .
\end{align}

%%%%%%%%%%%%%%%%%%
\begin{figure}[tbp]
\begin{center}
\includegraphics[height=2.4cm]{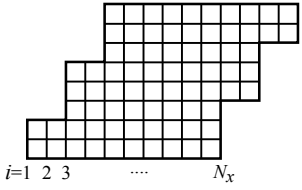}
\end{center}
\caption{Cross section of the square lattice used to apply
$H_{3 \rm D}$ in the case of $(9 \to 3/3/3)$ with $w = 2$.
}
\end{figure}
%%%%%%%%%%%%%%%%%%
Using $H_{2\rm D}$ and $H_{3\rm D}$, we determine the subband structure of
Dirac electrons in the cases of $(9 \to 3/3/3)$ and $(8 \to 3/2/3)$
for $w = 0$, $1$, and $2$.
To apply $H_{3\rm D}$ in determining the subband structure,
we set $B/A = 1.0$, $m_0/A = -0.5$, $m_{2\parallel}/A = 0.5$,
and $m_{2\perp}/A = -0.1$ and adopt the square lattice
with the cross section shown in Fig.~3 with $N_{x} = 25$.
Note that $H_{3\rm D}$ provides us with not only the subband structure of
Dirac electrons but also that of bulk electrons.
The factor $\eta (w)$ for $H_{2\rm D}$ is obtained as
$\eta(1) = 0.3$ and $\eta(2) = 0.09$ for the parameters given above.
This implies that the penetration depth of 1D helical edge modes of QSHIs
is on the order of one lattice constant.
The subband structures in the case of $(9 \to 3/3/3)$ are shown in Fig.~4.
In this case, one nondegenerate subband remains gapless as $M$ is odd
and the lowest gapped subband is doubly degenerate.
The excitation gap of the latter is reduced with increasing $w$
as $\Delta E/A = 0.618$ for $w = 0$,
$0.210$ for $w = 1$, and $0.064$ for $w = 2$.
As it is significantly reduced with increasing $w$,
we consider that the corresponding modes become pseudo-helical.
The subband structures in the case of $(8 \to 3/2/3)$ are shown in Fig.~5.
In this case, no exactly gapless subband appears as $M$ is even
and the lowest gapped subband is doubly degenerate.
The corresponding excitation gap $\Delta E/A = 0.347$ for $w = 0$
is markedly reduced with increasing $w$ as
$0.043$ for $w = 1$, and $0.004$ for $w = 2$,
leading to the appearance of two pseudo-helical modes.
In all the figures, the results obtained using $H_{2\rm D}$ and
$H_{3\rm D}$ are almost identical as far as lower subbands are concerned.
%%%%%%%%%%%%%%%%%%
\begin{figure}[tbp]
\begin{center}
\includegraphics[height=3.6cm]{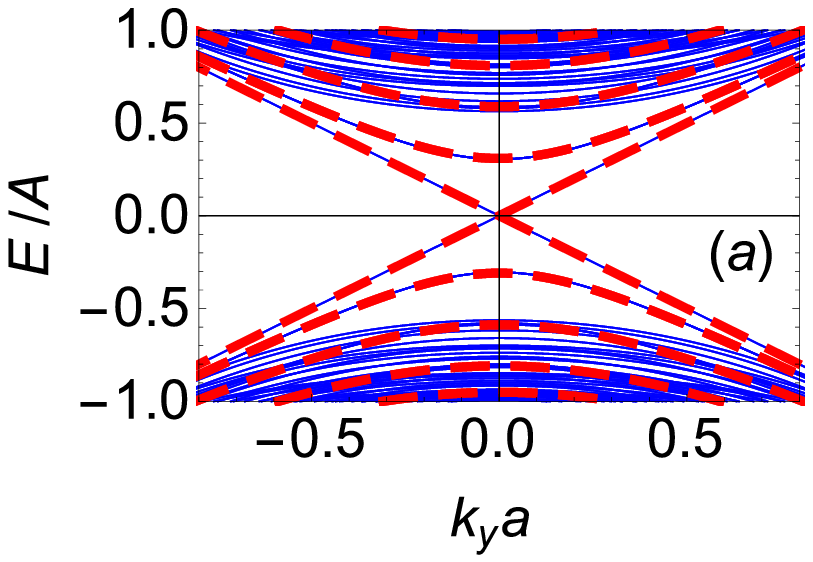}
\includegraphics[height=3.6cm]{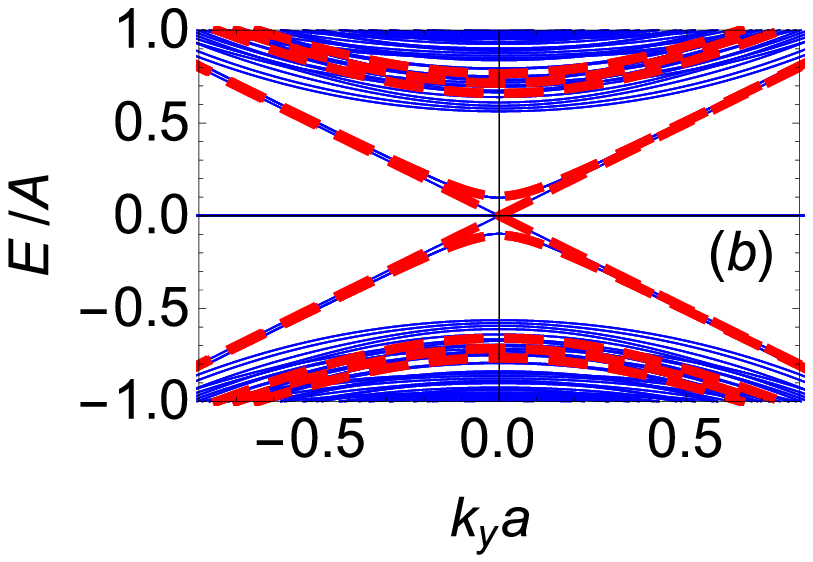}
\includegraphics[height=3.6cm]{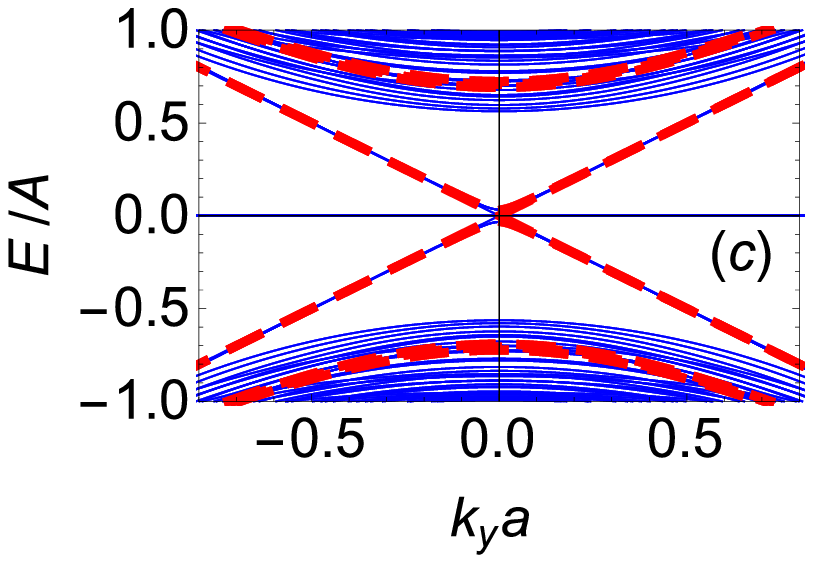}
\end{center}
\caption{(Color online) Subband structures in the case of $(9 \to 3/3/3)$
with (a) $w = 0$, (b) $w = 1$, and (c) $w = 2$,
where thick broken lines and thin solid lines respectively represent
the results obtained using $H_{2\rm D}$ and $H_{3\rm D}$.
}
\end{figure}
%%%%%%%%%%%%%%%%%%
%%%%%%%%%%%%%%%%%%
\begin{figure}[tbp]
\begin{center}
\includegraphics[height=3.6cm]{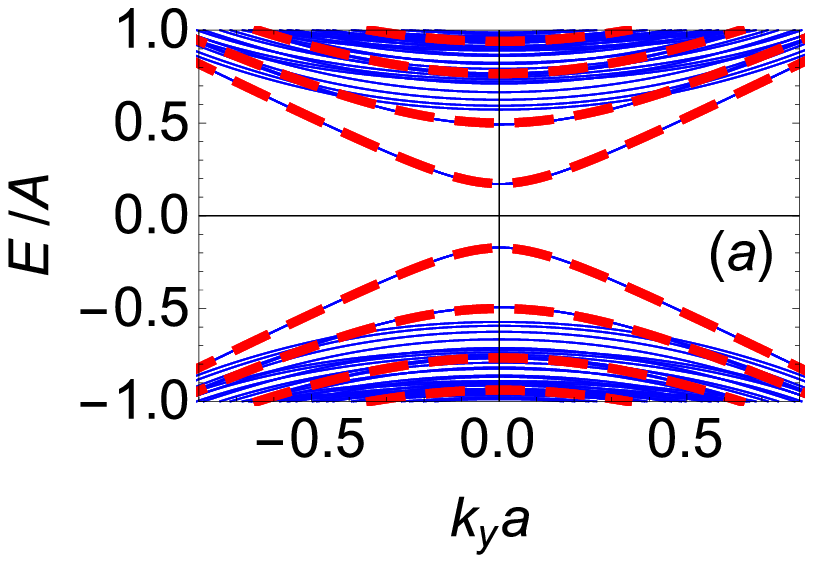}
\includegraphics[height=3.6cm]{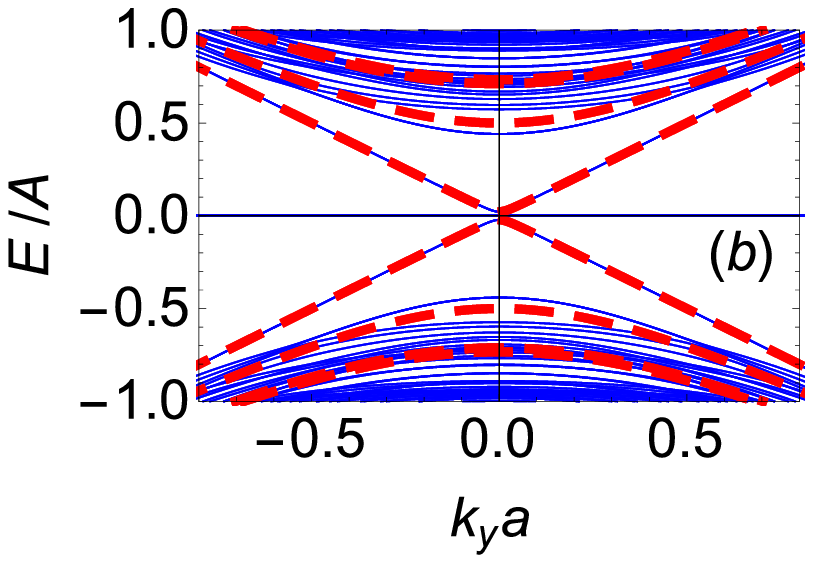}
\includegraphics[height=3.6cm]{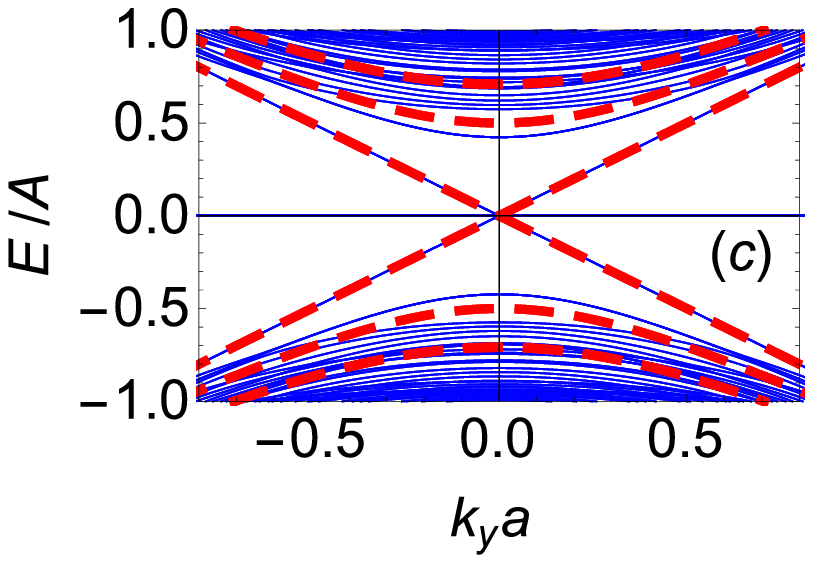}
\end{center}
\caption{(Color online) Subband structures in the case of $(8 \to 3/2/3)$
with (a) $w = 0$, (b) $w = 1$, and (c) $w = 2$,
where thick broken lines and thin solid lines respectively represent
the results obtained using $H_{2\rm D}$ and $H_{3\rm D}$.
}
\end{figure}
%%%%%%%%%%%%%%%%%%

Now, we consider the dimensionless conductance $g$ of
the Dirac electron system on a side surface of weakly disordered WTIs
in the cases of $(9 \to 3/3/3)$ and $(8 \to 3/2/3)$.
Following Ref.~\citen{takane1},
we calculate $g$ at zero temperature by using the effective model given in
Eq.~(\ref{H_2D}) with the replacement of $k_{y}$ with $-i\partial_{y}$.
We randomly distribute $\delta$-function-type impurities
in a finite region of $L > y > 0$ with $L = Na$ and
regard the semi-infinite regions of $y \ge L$ and $y \le 0$ as perfect leads.
The impurity potential $V$ is given by
$V = \sum_{j=1}^{M} V_{\rm imp}(y,j)|j\rangle \langle j|$ with
\begin{align}
  V_{\rm imp}(y,j)
  = \sum_{p} V_{p}a\delta(y-y_{p})\delta_{j,j_{p}} ,
\end{align}
where $V_p$ is the strength of the $p$th impurity located at $y = y_p$
on the $j_p$th chain.
We assume that $V_p$ is uniformly distributed within the interval of
$[-V_{0},+V_{0}]$.
To obtain $g$, we calculate the transmission matrix $\mib{t}$
through the disordered region, in terms of which
$g$ is determined as $g = {\rm tr}\{\mib{t}^{\dagger}\mib{t}\}$.
For a given impurity configuration, $\mib{t}$ is numerically obtained
employing the method presented in Ref.~\citen{tamura}.
To save computational time, we randomly choose $N$ points on the $y$-axis
within the sample region as $0<y_1<y_2\dots<y_N<L$ and then put an impurity
on every chain at each $y_l$ ($l = 1,2,\dots,N$).~\cite{takane1}
The total number of impurities is $M \times N$.
We compute the average dimensionless conductance $\langle g \rangle$
by setting $V_{0}/A = 0.5$~\cite{comment2} and placing
the Fermi level at $E_{\rm F}/A = 0.582$ and $0.35$
in the cases of $(9 \to 3/3/3)$ and $(8 \to 3/2/3)$, respectively.
In this setup, the gapless subband and the lowest gapped subband contribute
to electron transport in the case of $(9 \to 3/3/3)$,
while only the lowest gapped subband does so in the case of $(8 \to 3/2/3)$.
The number of samples, $N_{\rm sam}$, used in the average at each data point
is more than $2000$.
The relative uncertainty $\Delta g/\langle g \rangle$ with
$\Delta g \equiv ({\rm var}\{g\}/N_{\rm sam})^{1/2}$
is smaller than $0.012$.

The average dimensionless conductance in the case of $(9 \to 3/3/3)$ is
shown in Fig.~6.
In this case, $\langle g \rangle$ decreases to unity with increasing $L/a$
owing to the presence of a PCC.
We see that $\langle g \rangle$ is enhanced
with increasing width $w$ of steps.
This should be attributed to the contribution of pseudo-helical modes
induced by the step edges.
With the increase in $w$, their excitation gap $\Delta E$ decreases
and therefore their backscattering is suppressed.
This directly results in the enhancement of $\langle g \rangle$.
%%%%%%%%%%%%%%%%%%
\begin{figure}[tbp]
\begin{center}
\includegraphics[height=4.5cm]{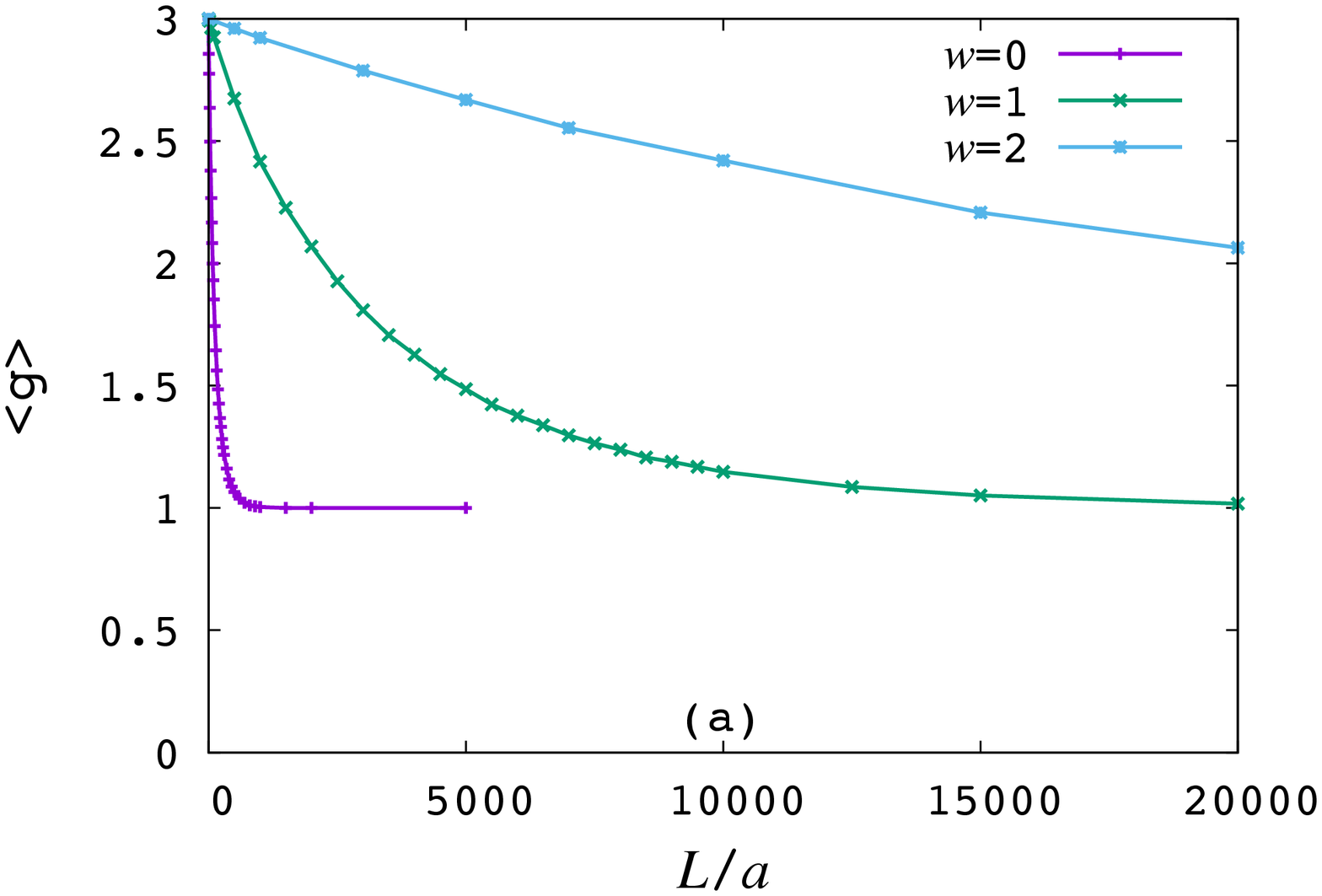}
\includegraphics[height=4.5cm]{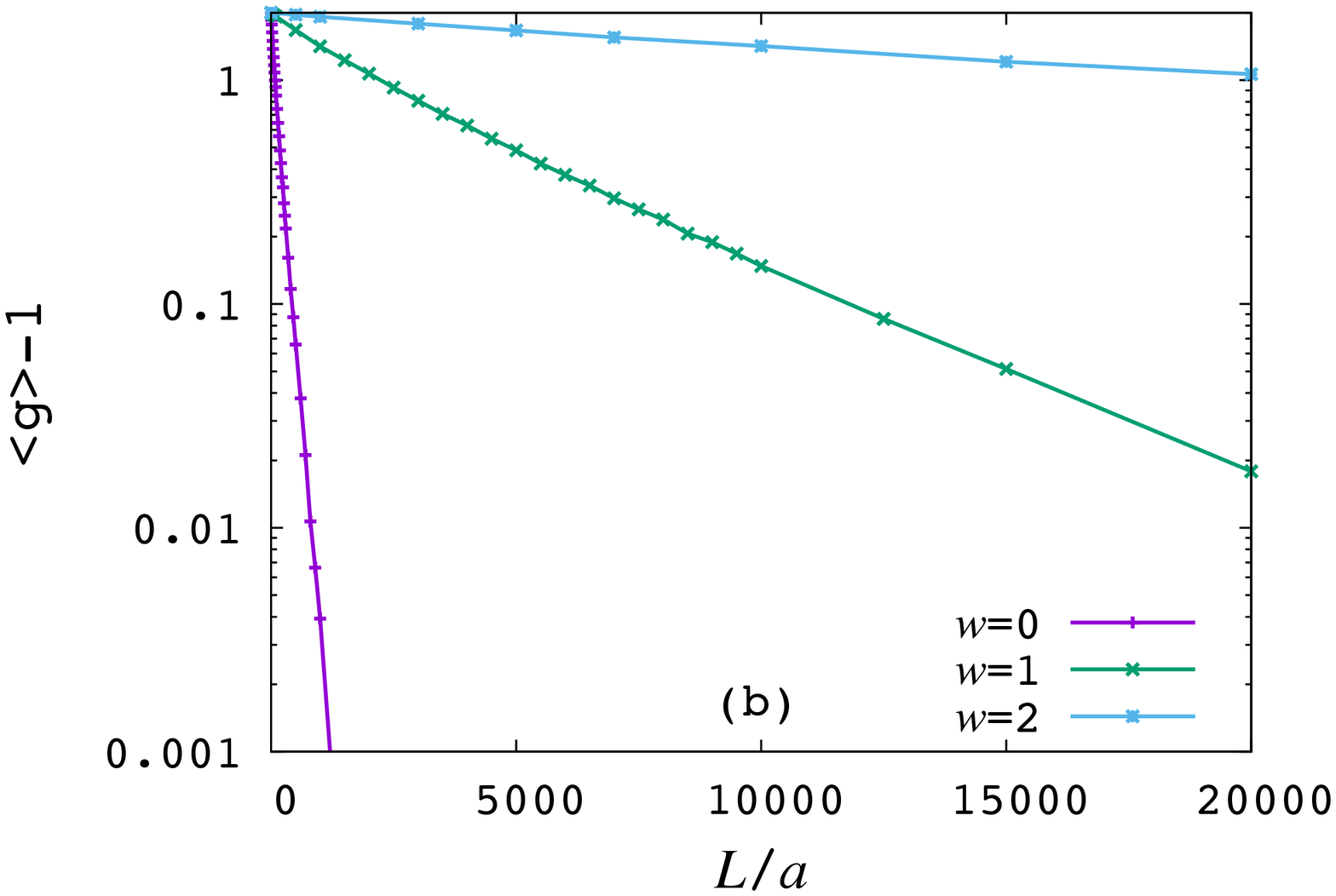}
\end{center}
\caption{(Color online) Average dimensionless conductance
as a function of $L/a$ in the case of $(9 \to 3/3/3)$
in (a) linear and (b) logarithmic scale,
where $\langle g \rangle -1$ is displayed in (b).
}
\end{figure}
%%%%%%%%%%%%%%%%%%
The average dimensionless conductance in the case of $(8 \to 3/2/3)$ is
shown in Fig.~7.
In this case, $\langle g \rangle$ decreases to zero with increasing $L/a$
reflecting the absence of a PCC.
We see that $\langle g \rangle$ is markedly enhanced with increasing $w$.
Particularly, $\langle g \rangle$ with $w = 2$ shows almost no decrease
within the interval of $L/a$ shown in Fig.~7.
This indicates that pseudo-helical modes become almost ideal in this case.
%%%%%%%%%%%%%%%%%%
\begin{figure}[tbp]
\begin{center}
\includegraphics[height=4.5cm]{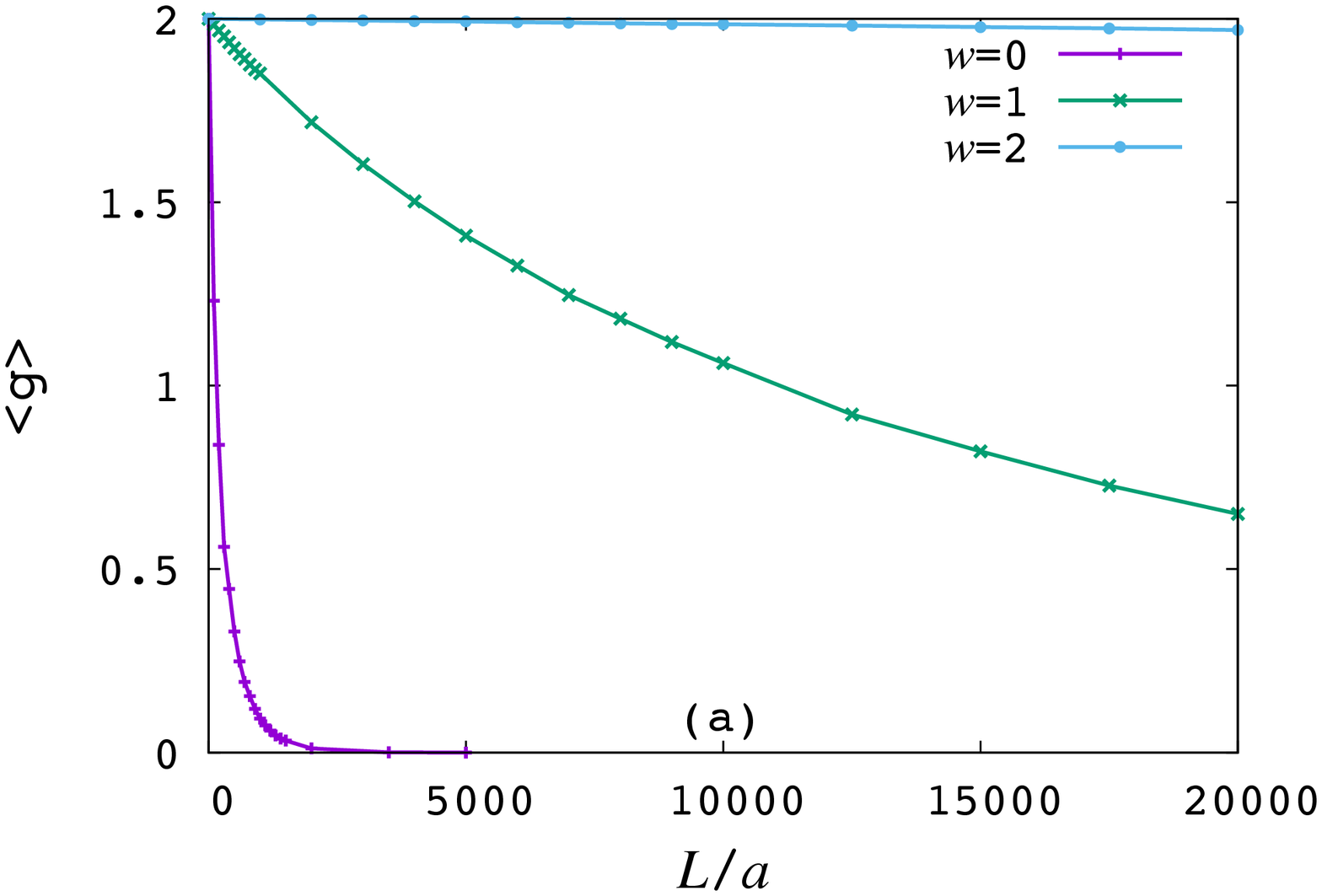}
\includegraphics[height=4.5cm]{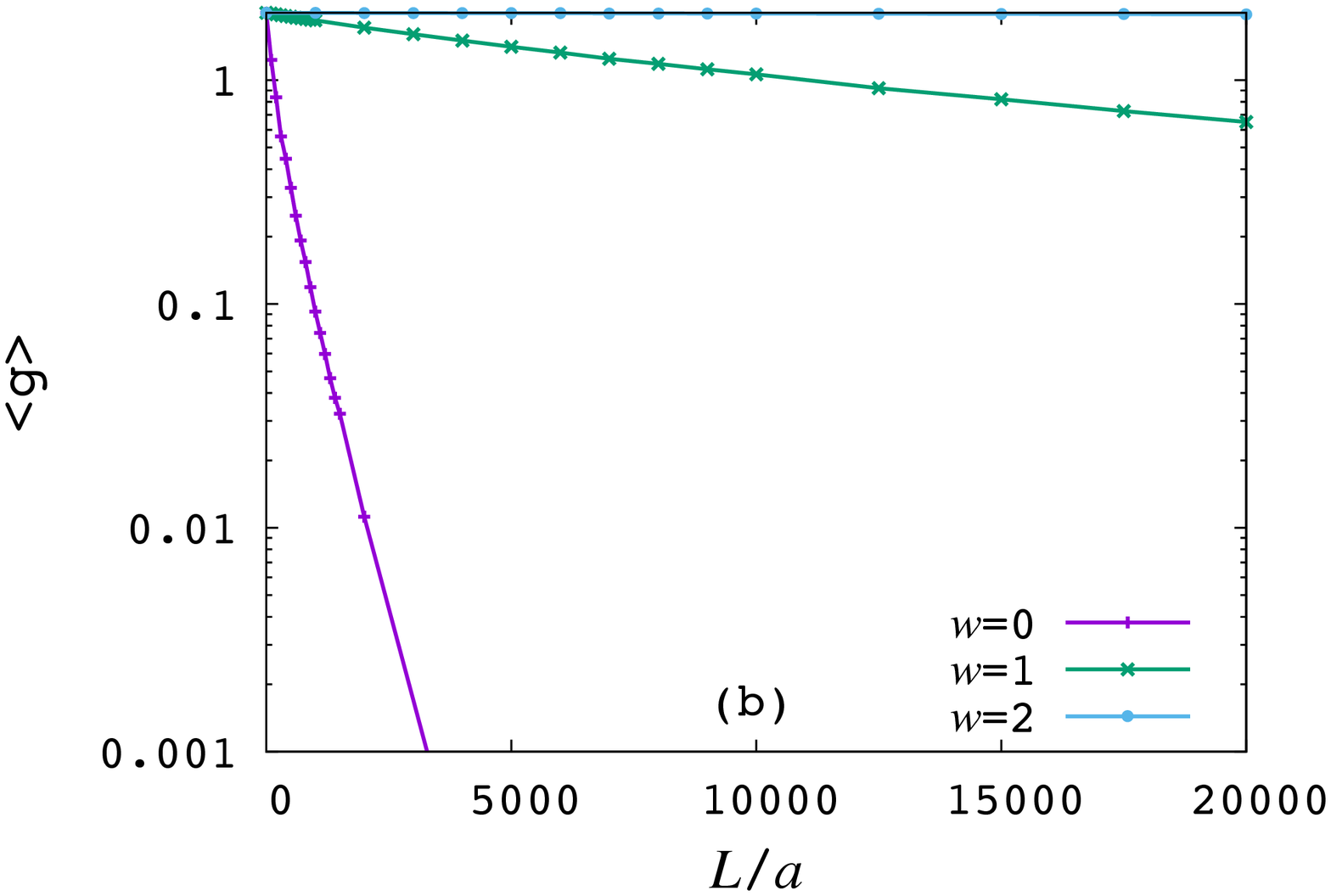}
\end{center}
\caption{(Color online) Average dimensionless conductance
as a function of $L/a$ in the case of $(8 \to 3/2/3)$
in (a) linear and (b) logarithmic scale.
}
\end{figure}
%%%%%%%%%%%%%%%%%%
We observe from Figs.~6 and 7 that the enhancement of $\langle g \rangle$ is
stronger in the case of $(8 \to 3/2/3)$ than in the case of $(9 \to 3/3/3)$.
This indicates that an even number of layers placed in between
pseudo-helical modes is very useful for enhancing the conductance of the system.

The above argument relies on $H_{2\rm D}$ for clearly observing
the role of pseudo-helical modes.
In other words, bulk states are ignored in calculating $\langle g \rangle$.
However, their effect can be neglected as long as the Fermi level
is much smaller than the energy gap of bulk states.~\cite{comment3}

In summary, we have shown that pseudo-helical modes
with a nearly gapless linear dispersion can be created on a side surface
of weak topological insulators by arranging straight step edges
following the proposed prescriptions.
We have also shown that such pseudo-helical modes significantly enhance
the conductance of the system.
Our proposal can be used as a useful building block in nanomaterials design.
Its experimental application may not be easy at present
but will be realized using the technique reported in Ref.~\citen{pauly}.
In realistic systems, the key parameter is
the penetration depth of one-dimensional helical edge modes.
We expect that the proposal becomes effective
in realistic weak topological insulators when the width of step edges
is comparable to, or larger than, the penetration depth.

\section*{Acknowledgment}

This work was supported by JSPS KAKENHI Grant Number 15K05130.


\begin{thebibliography}{99}

\bibitem{fu} L. Fu, C. L. Kane, and E. J. Mele,
Phys. Rev. Lett. {\bf 98}, 106803 (2007).

\bibitem{moore} J. E. Moore and L. Balents,
Phys. Rev. B {\bf 75}, 121306 (2007).

\bibitem{roy} R. Roy, Phys. Rev. B {\bf 79}, 195322 (2009).

\bibitem{kane} C. L. Kane and E. J. Mele,
Phys. Rev. Lett. {\bf 95}, 146802 (2005).

\bibitem{bernevig} B. A. Bernevig and S.-C. Zhang,
Phys. Rev. Lett. {\bf 96}, 106802 (2006).

\bibitem{ran} Y. Ran, Y. Zhang, and A. Vishwanath,
Nat. Phys. {\bf 5}, 298 (2009).

\bibitem{imura1} K.-I. Imura, Y. Takane, and A. Tanaka,
Phys. Rev. B {\bf 84}, 035443 (2011).

\bibitem{ringel} Z. Ringel, Y. E. Kraus, and A. Stern,
Phys. Rev. B {\bf 86}, 045102 (2012).

\bibitem{mong} R. S. K. Mong, J. H. Bardarson, and J. E. Moore,
Phys. Rev. Lett. {\bf 108}, 076804 (2012).

\bibitem{liu1} C.-X. Liu, X.-L. Qi, and S.-C. Zhang,
Physica E {\bf 44}, 906 (2012).

\bibitem{imura2} K.-I. Imura, M. Okamoto, Y. Yoshimura, Y. Takane,
and T. Ohtsuki, Phys. Rev. B {\bf 86}, 245436 (2012).

\bibitem{yoshimura} Y. Yoshimura, A. Matsumoto, Y. Takane, and K.-I. Imura,
Phys. Rev. B {\bf 88}, 045408 (2013).

\bibitem{kobayashi1} K. Kobayashi, T. Ohtsuki, and K.-I. Imura,
Phys. Rev. Lett. {\bf 110}, 236803 (2013).

\bibitem{morimoto} T. Morimoto and A. Furusaki,
Phys. Rev. B {\bf 89}, 035117 (2014).

\bibitem{obuse} H. Obuse, S. Ryu, A. Furusaki, and C. Mudry,
Phys. Rev. B {\bf 89}, 155315 (2014).

\bibitem{takane1} Y. Takane, J. Phys. Soc. Jpn. {\bf 83}, 103706 (2014).

\bibitem{arita} T. Arita and Y. Takane,
J. Phys. Soc. Jpn. {\bf 83}, 124716 (2014).

\bibitem{takane2} Y. Takane, J. Phys. Soc. Jpn. {\bf 84}, 084710 (2015).

\bibitem{matsumoto} A. Matsumoto, T. Arita, Y. Takane, Y. Yoshimura,
and K.-I. Imura, Phys. Rev. B {\bf 92}, 195424 (2015).

\bibitem{kobayashi2} K. Kobayashi, Y. Yoshimura, K.-I. Imura, and T. Ohtsuki,
Phys. Rev. B {\bf 92}, 235407 (2015).

\bibitem{yan} B.-H. Yan, L. M\"{u}chler, and C. Felser,
Phys. Rev. Lett. {\bf 109}, 116406 (2012).

\bibitem{rasche} B. Rasche, A. Isaeva, M. Ruck, S. Borisenko, V. Zabolotnyy,
B. Buchner, K. Koepernik, C. Ortix, M. Richter, and J. van den Brink,
Nat. Mater. {\bf 12}, 422 (2013). 

\bibitem{tang} P. Tang, B. Yan, W. Cao, S.-C. Wu, C. Felser, and W. Duan,
Phys. Rev. B {\bf 89}, 041409 (2014).

\bibitem{g-yang} G. Yang, J. Liu, L. Fu, W. Duan, and C. Liu,
Phys. Rev. B {\bf 89}, 085312 (2014). 

\bibitem{pauly} C. Pauly, B. Rasche, K. Koepernik, M. Liebmann, M. Pratzer,
M. Richter, J. Kellner, M. Eschbach, B. Kaufmann, L. Plucinski,
C. M. Schneider, M. Ruck, J. van den Brink, and M. Morgenstern,
Nat. Phys. {\bf 11}, 338 (2015).

\bibitem{zhou} J.-J. Zhou, W. Feng, G.-B. Liu, Y. Yao,
New J. Phys. {\bf 17}, 015004 (2015).

\bibitem{ando1} T. Ando and H. Suzuura,
J. Phys. Soc. Jpn. {\bf 71}, 2753 (2002).

\bibitem{takane3} Y. Takane,
J. Phys. Soc. Jpn. {\bf 73}, 9 (2004).

\bibitem{takane4} Y. Takane,
J. Phys. Soc. Jpn. {\bf 73}, 1430 (2004).

\bibitem{takane5} Y. Takane,
J. Phys. Soc. Jpn. {\bf 73}, 2366 (2004).

\bibitem{ando2} T. Ando,
J. Phys. Soc. Jpn. {\bf 75}, 054701 (2006).

\bibitem{liu2} C.-X. Liu, X.-L. Qi, H. Zhang, X. Dai, Z. Fang,
and S.-C. Zhang, Phys. Rev. B {\bf 82}, 045122 (2010).

\bibitem{comment1} In Ref.~\citen{arita},
$\eta(w)$ is determined by a variational calculation.
Equation~(\ref{eq:def-eta}) approximately reproduces the resulting
$\eta(w)$ in a simpler manner.

\bibitem{tamura} H. Tamura and T. Ando, Phys. Rev. B {\bf 44}, 1792 (1991).

\bibitem{comment2} As $g$ should obey
the scaling theory,~\cite{takane3,macedo} we expect that a variation
in $V_{0}$ results in only a renormalization of the mean free path.
Hence, the $L/a$ dependence of $\langle g \rangle$
is not qualitatively modified by such a variation.

\bibitem{macedo} A. M. S. Mac\^{e}do and J. T. Chalker,
Phys. Rev. B {\bf 46}, 14985 (1992).

\bibitem{comment3} Although the Fermi level assumed in our numerical
calculations does not satisfy this condition,
there is no essential difficulty in satisfying this in actual situations.




\end{thebibliography}
\end{document}